\documentclass[aps,amsmath,amssymb,prd,preprintnumbers,nofootinbib,twocolumn]{revtex4}

\usepackage{amsmath}
\usepackage{bm}
\usepackage[dvips]{graphicx}

\begin{document}

\title{A diagrammatic derivation of the meson effective masses in the neutral
color-flavor-locked phase of Quantum Chromodynamics}

\author{Marco~Ruggieri}\email{marco.ruggieri@ba.infn.it}
\affiliation{Dipartimento di Fisica, Universit\`a di Bari, I-70126 Bari,
Italia} \affiliation{I.N.F.N., Sezione di Bari, I-70126 Bari,
Italia}

\preprint{BARI-TH 565/07}

\begin{abstract}We offer a diagrammatic derivation of the effective
masses of the axial flavor excitations in the electrical and color
neutral CFL phase of QCD. In particular we concentrate on the
excitations with the quantum numbers of the kaons: we show how their
effective chemical potentials, responsible of their Bose-Einstein
condensation and found previously on the basis of pure symmetry
arguments, arise at the microscopic level by loop effects. We
perform also the numerical evaluation of the relevant loops in the
whole CFL regime $M_s^2/2\mu\Delta\leqslant 1$, showing the
existence of the enhancement of the kaon condensation with respect
to the lowest order result. Finally we discuss the role of color
neutrality in the microscopic calculation.\end{abstract}


\maketitle

\section{Introduction}
The study of the phase diagram of Quantum Chromodynamics (QCD) has
attracted a lot of interest in the last years. In particular, the
major theoretical improvement of the knowledge of the high density
and low temperature regime of QCD has been achieved by
phenomenological studies based both on Nambu-Jona Lasinio (NJL)
models~\cite{Abuki:2004zk,Ruster:2005jc,Blaschke:2005uj} and
abstract formulations relying on Ginzburg-Landau expansions of the
free energy~\cite{Hatsuda:2006ps,Yamamoto:2007ah}. In these extreme
conditions the QCD ground state is likely to be a color
superconductor (for reviews
see~\cite{Rajagopal:2000wf,Alford:2001dt,Nardulli:2002ma,
Reddy:2002ri,Schafer:2003vz,Rischke:2003mt,Buballa:2003qv,
Huang:2004ik,Shovkovy:2004me,Alford:2006fw}).

At asymptotic densities (or equivalently, at asymptotic baryon
chemical potential $\mu$) and low temperatures the ground state of
three flavor quark matter is the color-flavor-locked state
(CFL)~\cite{Alford:1998mk,Evans:1999at}, where the color and flavor
symmetries of QCD are broken spontaneously by a di-quark vacuum
expectation value proportional to the gap parameter $\Delta$, but a
residual group linking color and vector flavor symmetries is left.
In the CFL phase all the quarks pair among themselves and get an
effective mass $\Delta$.

When the density is decreased, finite quark masses effects and
electrical and color neutrality conditions break the symmetric CFL
state into a less symmetric state. Examples include the gapless CFL
state~\cite{Alford:2003fq}, kaon condensed
states~\cite{Bedaque:2001je,Kaplan:2001qk}, inhomogeneous three
flavor color superconductors~\cite{Casalbuoni:2005zp}, gapless two
flavor color superconductors, both in the
homogeneous~\cite{Shovkovy:2003uu,Gubankova:2003uj} and
inhomogeneous realizations~\cite{Alford:2000ze}, and spin one
pairing~\cite{Schafer:2000tw,Aguilera:2005tg}. In particular,
denoting by $M_s$ the in-medium strange quark mass, we remind that
the parameter useful to describe the CFL$\rightarrow$gCFL transition
is $M_s^2/2\mu\Delta$: if it is smaller than 1 then the ground state
is the CFL phase; on the other hand, if $M_s^2/2\mu\Delta > 1$ then
the ground state is the gapless gCFL phase, characterized by gapless
quark excitations in the spectrum. Unfortunately gapless homogeneous
phases suffer the problem of Meissner anti-screening: the squared
Meissner masses of some of the gluons, when evaluated in the one
loop approximation, turn out to be negative, both in the two
flavor~\cite{Huang:2004bg} and in the three
flavor~\cite{Casalbuoni:2004tb} case. This fact is interpreted as an
instability toward a gluon condensed state~\cite{Gorbar:2005rx}
and/or an inhomogeneous ground states, with meson currents
spontaneously
generated~\cite{Huang:2005pv,Gatto:2007ja,Schafer:2005ym} and non
isotropic pairing~\cite{Giannakis:2004pf} among quarks.

In the CFL state the chiral symmetry is broken, and an octet of
pseudoscalar meson fields appear in the spectrum as required by the
Goldstone theorem. Beside them two flavor singlet fields arise
because of the breaking of the baryon $U(1)_V$ and of the axial
$U(1)_A$ which is restored at high densities.

The fact that all the quarks are gapped implies the existence of two
characteristic energy scales in the CFL phase: the gap parameter
$\Delta$, that roughly speaking measures the mass of the fermion
excitations, and the high baryon chemical potential $\mu \gg
\Delta$. Since the gap is much smaller than the chemical potential
(and thus of the Fermi momenta of the quarks) the quark dynamics is
dominated by the fermion modes living in a thin shell around the
Fermi surface: one can therefore write a renormalized effective
fermion action (in the sense of Wilson), in terms of the soft modes
that live near the Fermi
surface~\cite{Nardulli:2002ma,Hong:1998tn,Schafer:2003jn}. Moreover,
since the fermion excitations are gapped, for energies much smaller
than the gap one can describe the low energy physics of the CFL
phase by means of the light flavor
excitations~\cite{Casalbuoni:1999wu,Bedaque:2001je}.

It was noticed~\cite{Bedaque:2001je} that a finite strange quark
mass can modify the CFL ground state before the transition to gCFL
occurs: as a matter of fact, in the limit of isospin symmetry $M_u =
M_d$ (capital letters denote in-medium quark masses), the
excitations with kaon quantum numbers feel an effective chemical
potential that can become larger than the kaon mass if the strange
quark mass is large enough. If this is the case then it is possible
to lower the free energy of the system by forming a kaon
Bose-Einstein condensate. Astrophysical implications of the presence
of a kaon condensed phase in the core of a compact star have been
widely studied~\cite{Kaplan:2001hh}. It becomes clear then that the
knowledge of the effective masses of the kaon modes in the CFL phase
is important for the determination of the true ground state of three
flavor QCD.

The important project of writing an effective lagrangian for the
flavor excitations, and thus of their effective masses, in the CFL
phase of QCD is a classical topic widely
covered~\cite{Son:1999cm,Casalbuoni:2000na,Beane:2000ms,Schafer:2001za,Manuel:2000wm};
it has been considered more recently
in~\cite{Kryjevski:2003cu,Buballa:2004sx,Forbes:2004ww,Ebert:2006tc},
where the effects of electrical and color neutrality and of the
finite quark masses have been introduced. However, we have noticed
that up to now there exists no clear derivation of the effective
masses of the kaons in the CFL phase, and in particular of their
effective chemical potentials, based on a purely microscopic model
of quarks interacting with the flavor excitations. We wish to cover
this important topic here.

The main scope of our work is to show how the effective chemical
potentials felt by the kaon fields, responsible of the kaon
condensation phenomenon and deduced on the basis of symmetries
in~\cite{Bedaque:2001je}, arise as loop effects once the octet of
the $SU(3)_A$ flavor excitations is introduced as an external
field~\cite{Casalbuoni:2000na,Eguchi:1976iz}. Recently the same
procedure has been applied in Ref.~\cite{Anglani:2007aa} to the
Goldstone modes of the inhomogeneous three flavor superconductive
state.

From the very beginning we work in an electrically and color neutral
state. This is the case because any homogeneous ground states in QCD
has to be color
neutral~\cite{Kryjevski:2003cu,Forbes:2004ww,Alford:2002kj}.
Moreover, one can show, by using a low energy effective lagrangian
of the CFL
phase~\cite{Casalbuoni:1999wu,Kryjevski:2003cu,Forbes:2004ww} that
color neutrality implies the decoupling, from the low energy
spectrum, of the colored degrees of freedom, leaving a lagrangian
written in terms of the only color singlet axial flavor excitations
(we neglect the flavor singlets corresponding to $U(1)_B$ and
$U(1)_A$). This turns to be equivalent to impose color neutrality in
the microscopic model~\cite{Kryjevski:2003cu,Forbes:2004ww}: as a
consequence, in order to reproduce properly such a low energy
effective lagrangian by a microscopic theory of interacting quarks,
the starting point must be a color neutral state.

We perform the calculation of the effective meson chemical
potentials both analytically, by an expansion in powers of
$M_s^2/\mu\Delta$, and numerically. The method presented here is
applicable to the gapless CFL phase as well. The analytical
calculation shows an interesting aspect: if one does not forces
color neutrality, and leaves the color chemical potentials
arbitrary, then one notices that the effective mass of the kaons
does not recover the value obtained by the effective theory. Only if
the color chemical potentials are chosen in order to fulfill the
neutrality condition, that is only if one considers a neutral state,
then the two results match.

The paper is organized as follows: in Section~\ref{sec:quarks} we
describe the quark content of the model. In Section~\ref{sec:R} we
introduce the $SU(3)_A$ flavor excitations as external fields via
the Eguchi recipe~\cite{Eguchi:1976iz}, defining their coupling to
the quarks and their one loop effective Lagrangian by integrating
over the quarks in the functional integral. Section~\ref{sec:Analyt}
is the main body of the paper: we compute both analytically and
numerically the effective chemical potentials felt by the flavored
meson fields, once the chemical potentials of the quarks are chosen
in order to fulfill the neutrality conditions (since we work in the
limit $M_u = M_d = 0$ only the kaons are massive, while all of the
other meson modes are
massless~\cite{Beane:2000ms,Bedaque:2001je,Schafer:2001za}).
Finally, we draw our conclusions, indicating some consequences of
our results and possible improvements of the work.

\section{The effective quark lagrangian}\label{sec:quarks}
In this paper we deal with three flavor quark matter, whose
interaction is modeled by a local Nambu-Jona Lasinio (NJL)
lagrangian~\cite{Nambu:1961tp} (for reviews
see~\cite{Buballa:2003qv,Klevansky:1992qe,Hatsuda:1994pi}). The main
limitation of NJL models is the lack of gluons; nevertheless, as it
is consistent with the global symmetries of QCD, it is believed to
be able to capture the essential physics of the problem.

At finite chemical potential and in presence of color condensation
the quark lagrangian is given by
\begin{equation}
{\cal L} = \bar\psi\left(i\partial_\mu \gamma^\mu +
\hat\mu\gamma_0\right)\psi - M_f \bar{\psi}_f \psi_f +{\cal
L}_\Delta~.\label{eq:lagr1MU}
\end{equation}
In the above equation $\hat\mu$ is the quark chemical potential
matrix, with color and flavor indices. It depends on $\mu$ (the
average quark chemical potential), $\mu_e$ (the electron chemical
potential), and $\mu_3,\,\mu_8$ (color chemical potentials)
\cite{Alford:2003fq}. For color and electric neutrality to be
implemented it is sufficient to consider only these chemical
potentials, related as they are to the charge matrix and the
diagonal color operators $T_3 = \frac 1 2 {\rm diag}(1,-1,0)$ and
$T_8 = \frac{1}{2 \sqrt 3 }{\rm diag}(1,1,-2)$ (in general one
should introduce a color chemical potential for each $SU(3)$ color
charge; however, as shown in~\cite{Buballa:2005bv}, for the
condensate with the color-flavor structure considered in this paper
it is enough to consider only $\mu_3$ and $\mu_8$, since the charges
related to the other color generators automatically vanish).
Therefore the matrix $\hat\mu$ is written as follows
\begin{eqnarray}
&&{\hat\mu}_{ij}^{\alpha\beta} = \left(\mu \delta_{ij} - \mu_e
Q_{ij}\right)\delta^{\alpha\beta}+ \delta_{ij} \left(\mu_3
T_3^{\alpha\beta}+\frac{2}{\sqrt 3}\mu_8
T_8^{\alpha\beta}\right)\nonumber\\ \label{9}
\end{eqnarray} with $Q= {\rm diag} (2/3,-1/3,-1/3)$ ($i,j
=1,3 $ flavor indices; $\alpha,\beta =1,3 $ colour indices).

The term ${\cal L}_\Delta$ is responsible for color condensation,
and is given in the mean field approximation by
\begin{equation}
{\cal L}_\Delta = -\frac{1}{2}\sum_{I=1}^3\left(\Delta_{I}({\bm
r})\psi_{\alpha i}^\dagger \gamma_5\epsilon^{\alpha\beta
I}\epsilon_{i j I} C \psi_{\beta j}^* + h.c. \right)
~.\label{eq:LagrDelta2}
\end{equation}
Eq.~\eqref{eq:LagrDelta2} describes the fact that in the ground
state one has a non-vanishing expectation value of the di-quark
field operator
\begin{equation}
\langle\psi({\bm r})_{\alpha i} \psi({\bm r})_{\beta j}\rangle
\propto \Delta_I({\bm r}) \epsilon_{\alpha \beta I}\epsilon_{ijI}
\neq 0~.\label{eq:SSB}
\end{equation}
In this work we are interested to the CFL
phase~\cite{Alford:1998mk,Alford:2003fq}: in this case $\Delta_1 =
\Delta_2 = \Delta_3 \equiv \Delta$, independent of the space
coordinates ${\bm r}$.

Finally, $M_f$ in Eq.~\eqref{eq:lagr1MU} denote the in-medium quark
mass of the flavor $f$. Here we treat the quark masses at the
leading order in $M_f^2/\mu$: at this order the effect of the finite
mass is a shift of the quark chemical potentials $\mu_f$ by the
amount $-M_f^2/2\mu$. This is a widely used approximation, which
captures the main role of the finite quark masses, namely the
reduction of their Fermi spheres (in addition to that, the
approximation allows for easier calculations).

In this paper we adopt the high density effective description of
QCD~\cite{Nardulli:2002ma,Hong:1998tn,Schafer:2003jn}: this
approximation amounts to consider only the quarks with momenta close
to the Fermi surface; this approximation is justified since in the
weak coupling regime, to which we are interested here, the quarks
living inside the Fermi sphere are Pauli blocked and are not
relevant for the dynamics; moreover, the negative energy fields give
rise to operators that are formally suppressed by inverse powers of
$\mu$ and therefore give a negligible contribution to the quark
propagator.

The high density effective lagrangian of the quarks in the CFL phase
of QCD has been discussed many times in the literature (see for
example~\cite{Nardulli:2002ma,Beane:2000ms,Schafer:2001za}),
therefore here we simply quote the result in the momentum space,
namely
\begin{eqnarray}
&&{\cal L}=\frac{1}{2}\int\!\frac{d{\bm n}}{4\pi}~\chi^\dagger_A
\left(\begin{array}{cc}
       K_{AB}(\ell)  & -\Delta_{AB} \\
        -\Delta^\star_{AB} & \tilde{K}_{AB}(\ell)
      \end{array}
\right)\chi_B ~ \nonumber\\
&&~~~~~~~~~~~~~~~~~~~~~+ L\rightarrow R~.\label{eq:Lagr1}
\end{eqnarray}
Here $A=1,\dots,9$ is a color-flavor index; the rotation to the new
basis is performed by means of the matrices $F_A$ defined
in~\cite{Casalbuoni:2004tb}. The kinetic terms are defined as
$K_{AB} = V\cdot\ell~\delta_{AB} + \delta\mu_{AB}$~, $\tilde{K}_{AB}
= \tilde{V}\cdot\ell~\delta_{AB} - \delta\mu_{AB}$. The quark
momenta are measured as ${\bm p} = \mu{\bm n}+{\bm \ell}$, $p_0 =
\ell_0$, with ${\bm n}$ a unit vector denoting the Fermi velocity of
the quarks and $\mu$ is a reference large momentum, usually equal to
the baryon chemical potential. The chemical potential of the quark
with index $A$ is written as $\mu_A = \mu + \delta\mu_A$ and
$\delta\mu_{AB} \equiv \delta\mu_A \delta_{AB}$. The entry
$\delta\mu_A$ contains also the effective mass $-M_f^2/2\mu$ of the
flavor $f$. The gap matrix is given by $\Delta_{AB} = \Delta_I({\bm
r})\text{Tr}[\epsilon_I F_A^T \epsilon_I F_B]$. Finally, we have
introduced the Nambu-Gorkov doublet for the left-handed fields,
\begin{equation}
\chi = \left(\begin{array}{c}
               \psi({\bm n}) \\
               C\psi^*(-{\bm n})
             \end{array}
\right)~.
\end{equation}

The calculations presented in the Section~\ref{sec:Analyt} are
devoted to the evaluation of the masses of the excitations related
to the breaking of the $SU(3)_A$ symmetry. In that context one needs
to add mass corrections to the high density effective theory (HDET).
One way to do that consistently in a NJL model is to introduce an
anti-gap term to the above lagrangian: it has been treated many
times in the literature, and its role in the calculation of
pseudo-Nambu-Goldstone modes in NJL studies of high density QCD has
been emphasized. In particular we need the lagrangian which
describes antiquark--antiquark pairing, introduced in the HDET
formalims in~\cite{Beane:2000ms,Casalbuoni:2002st}. On the other
hand, starting from the QCD quark-gluon vertex it is possible to get
effective four fermion interactions by integrating out the electric
gluons, see for example the clear discussion
in~\cite{Schafer:2001za}. The introduction of such vertices is
essential in QCD since it has been shown that the anti-gap
lagrangian gives rise to gauge dependent values of the shift in the
vacuum energy generated by the finite values of the quark
masses~\cite{Schafer:2001za}; also, when one considers
quark-antiquark pairing, the shift of the vacuum energy sum up to
zero and one is left with the quark-quark contribution only, see
Eq.~(39) of Ref.~\cite{Schafer:2001za}. In any case (NJL or QCD) the
contribution of this kind of corrections to the squared meson masses
is highly suppressed since it contains light quark mass insertions.
As a consequence one can obtain them in the microscopic calculation
by putting $M_s = 0$ in the quark propagators, and the well known
results hold~\cite{Son:1999cm,Casalbuoni:2000na,Beane:2000ms,
Schafer:2001za,Manuel:2000wm}.

We close this section by defining the left handed quark propagator
in momentum space,
\begin{equation}
S(\ell)^{-1}_{AB} = \left(\begin{array}{cc}
       K_{AB}(\ell)  & -\Delta_{AB} \\
        -\Delta^\star_{AB} & \tilde{K}_{AB}(\ell)
      \end{array}
\right)~;\label{eq:FermProp}
\end{equation}
the analogous Green function of the right handed quarks is obtained
trivially from the previous one. From now on we consider the limit
$M_u = M_d = 0$. Thus the electrical and color neutrality conditions
are fulfilled by~\cite{Alford:2003fq,Steiner:2002gx}
\begin{eqnarray}
&&\mu_e = \mu_3 = 0~, \label{eq:MU3}\\
&&\mu_8 \approx -\frac{M_s^2}{2\mu}~.\label{eq:MU8}
\end{eqnarray}
After these assumptions are made we write explicitly the chemical
potentials entering into the quark Lagrangian~\eqref{eq:Lagr1} in
the case of the neutral CFL phase: $\mu_A = \mu + \delta\mu_A$
with~\cite{Alford:2003fq}
\begin{eqnarray}
&&\delta\mu_{1} = \delta\mu_{ur} =\mu - \frac{M_s^2}{6\mu} ~,
\label{eq:1}\\
&&\delta\mu_{2} = \delta\mu_{dg} = \delta\mu_{1}~, \\
&&\delta\mu_{3} = \delta\mu_{bs} = \delta\mu_{1}~,
\end{eqnarray}
\begin{eqnarray}
&&\delta\mu_{4} = \delta\mu_{dr} = \delta\mu_{1}~, \\
&&\delta\mu_{5} = \delta\mu_{ug} = \delta\mu_{1}~,
\end{eqnarray}
\begin{eqnarray}
&&\delta\mu_{6} = \delta\mu_{sr} = \delta\mu_{1} -\frac{M_s^2}{2\mu} ~, \\
&&\delta\mu_{7} = \delta\mu_{ub} = \delta\mu_{1}
+\frac{M_s^2}{2\mu}~,
\end{eqnarray}
\begin{eqnarray}
&&\delta\mu_{8} = \delta\mu_{sg} = \delta\mu_{1} -\frac{M_s^2}{2\mu}~, \\
&&\delta\mu_{9} = \delta\mu_{db} = \delta\mu_{1}
+\frac{M_s^2}{2\mu}~.\label{eq:9}
\end{eqnarray}
From the above equations we notice that the chemical potentials of
the quarks with $A=1,\dots,5$ is the same: thus it results more
convenient, in the HDET momenta decomposition, to choose
$\mu+\delta\mu_1$ as the large reference momentum.

\section{Coupling of the $SU(3)_A$ Goldstone
bosons to the quarks}\label{sec:R} In this section we derive the
coupling of the quarks to the Goldstones. Following
Ref.~\cite{Eguchi:1976iz} the $SU(3)_A$ excitations are introduced
in the model by rotating the CFL quark condensate. This rotation can
be achieved on the left handed fields by means of the axial flavor
transformations defined by~\cite{Casalbuoni:2000na}
\begin{equation}
\psi_{\alpha i} \rightarrow \psi_{\alpha k}\left({\cal
U}^\dagger\right)_{ki}~,~~~~{\cal U} \equiv \exp
\displaystyle{\left\{i \frac{\pi_a
\lambda_a}{2F}\right\}}~,\label{eq:octet}
\end{equation}
where $a=1,...,8$, $\lambda_a$ are the Gell-Mann matrices,
normalized as $\text{Tr}\{\lambda_a \lambda_b\}=2 \delta_{ab}$~, and
$F$ is the decay constant. For the right fields the transformation
is analogous.

In order to properly describe the $SU(3)_A$ Goldstone excitations we
promote the quark mass matrix $M$ to a spurion field that has
definite transformations under chiral transformations, that is
\begin{equation}
M \rightarrow L M R^\dagger~.\label{eq:Mtransformation}
\end{equation}
An analogous transformation is introduced for the charge matrix:
$Q\rightarrow L Q L^\dagger$. In this way the effective quark mass
term for the left handed fields, namely $-\psi^\dagger (M
M^\dagger/2\mu) \psi$, and the charge chemical potential term
$-\psi^\dagger (\mu_e Q) \psi$ are invariant under the chiral
transformations (the same is true for the right handed fields), and
thus under the quark rotation defined by Eq.~\eqref{eq:octet}. The
lagrangian is thus given by
\begin{equation}
{\cal L} = \int\!\frac{d {\bm n}}{8\pi}\chi^\dagger_A
\left(\begin{array}{cc}
        K_{AB}(\ell) & -\Xi_{BA}^\star \\
        -\Xi_{AB} & \tilde{K}_{AB}(\ell)
      \end{array}
\right)\chi_B~,\label{eq:parapapapa}
\end{equation}
where
\begin{equation}
\Xi_{AB} =\Delta_I^\star({\bm r})\text{Tr}[\epsilon_I (F_A {\cal
U}^\dagger)^T \epsilon_I F_B {\cal U}^\dagger]~.
\end{equation}
After the rotation is done, one sets the chemical potentials and the
quark masses to their values in the neutral CFL phase. From the
above equation it is clear that the $SU(3)_A$ flavor excitations are
introduced only as a rotation of the quark condensate, and disappear
from the spectrum if at this stage one sets $\Delta = 0$.
\begin{widetext}
Eq.~\eqref{eq:parapapapa} is the non linear realization of a theory
of quarks interacting with an octet of external fields. In order to
obtain a kinetic term for the ``pions'' $\pi_a$ we linearize the
theory by expanding ${\cal U}$ in Eq.~\eqref{eq:octet} up to the
second order in the $\pi_a$ fields: this results in three-body and
four-body interaction interaction terms,
\begin{equation}
i{\cal L}_{\chi\chi\pi}= +\frac{i\,\pi_a}{2F} \int\!\frac{d {\bm
n}}{8\pi}\chi^\dagger_A ({\cal G}_3)_{AB}^a\,
\chi_B~,\label{eq:octetLagr3}
\end{equation}
\begin{equation}
i{\cal L}_{\chi\chi\pi\pi}= +\frac{\pi_a \pi_b}{8 F^2} \int\!\frac{d
{\bm n}}{8\pi}\chi^\dagger_A ({\cal G}_4)_{AB}^{ab}\,
\chi_B~.\label{eq:octetLagr4}
\end{equation}
The expressions of ${\cal G}_3,{\cal G}_4$ are as follows:
\begin{equation}
{\cal G}_3=\left(\begin{array}{cc}
    0     & - (K^{3a}_{BA})^\star \\
       K^{3a}_{AB} & 0
      \end{array}
\right)~,\label{eq:octetLagrAPP3}
\end{equation}
\begin{equation}
{\cal G}_4=\left(\begin{array}{cc}
        0 & (K^{4 ab}_{BA})^\star  \\
      K^{4 ab}_{AB} & 0
      \end{array}
\right)~.\label{eq:octetLagrAPP4}
\end{equation}
The off-diagonal entries are defined as
\begin{eqnarray}
K^{3a}_{AB} &=& \Delta_I^\star({\bm r}) ~\text{Tr}[\epsilon_I
\lambda_a^T F_A^T \epsilon_I F_B
+\epsilon_I F_A^T \epsilon_I F_B \lambda_a ]~,\label{eq:K3}\\
K^{4ab}_{AB} &=& \Delta_I^\star({\bm r}) ~\text{Tr}[\epsilon_I
\lambda_a^T \lambda_b^T F_A^T \epsilon_I F_B + \epsilon_I F_A^T
\epsilon_I F_B \lambda_a \lambda_b + 2 \epsilon_I \lambda_a^T F_A^T
\epsilon_I F_B \lambda_b]~.\label{eq:K4}
\end{eqnarray}

Integrating over the fermion fields in the generating functional of
the model one is left with the effective lagrangian in momentum
space ${\cal L}(p) = {\cal L}_{s.e.}(p) + {\cal L}_{tad}$
with~\cite{Nardulli:2002ma}
\begin{eqnarray}
i{\cal L}_{tad} &=& +\left(\frac{\pi_a \pi_b}{8
F^2}\right)\frac{\mu^2}{4\pi^3 } \int\!\frac{d{\bm n}}{4\pi}\int
d^2\ell~\text{Tr}[ S(\ell)  {\cal
G}^4]~,\label{eq:tadpoleMom} \\
i {\cal L}_{s.e.}(p) &=& -
\frac{1}{2}\left(i\frac{\pi_a}{2F}\right)\left(i\frac{\pi_b}{2F}\right)
\frac{\mu^2}{4\pi^3}\int\!\frac{d{\bm n}}{4\pi}\int
d^2\ell~\text{Tr}[ S(\ell + p)  {\cal G}^3  S(\ell)  {\cal
G}^3]~;\label{eq:selfMom}
\end{eqnarray}
the overall minus sign in Eq.~\eqref{eq:selfMom} is due to the
fermion loop. We have already kept into account of the $L+R$
contribution, and the trace is intended in Nambu-Gorkov as well as
in color-flavor indices. The quark propagator is defined in
Eq.~\eqref{eq:FermProp} with chemical potentials given in
Eqs.~\eqref{eq:1}~-~\eqref{eq:9}.
\end{widetext}

\section{Meson masses}\label{sec:Analyt}

The most important task in this work is the computation of the
squared masses of the axial flavor excitations, to which we turn. We
begin with the analytical calculation at small $M_s^2/2\mu\Delta$;
the analytical results are confirmed by a numerical analysis, that
allows to make evaluations up to $M_s^2/2\mu\Delta = 1$ (for higher
values of the ratio $M_s^2/2\mu\Delta$ one enters the gCFL regime:
in this case the role of gapless fermions has to be kept into
account, together with $\mu_e \neq 0$ and $\mu_3 \neq 0$).

\subsection{Analytical results}
In this section we present analytical results that can easily
obtained by expanding the quark propagator in powers of
$M_s^2/\mu\Delta$. The result of the calculation is in agreement
with the result of~\cite{Schafer:2001za,Bedaque:2001je}, obtained by
the authors on the basis of pure symmetry arguments. For simplicity,
since we are interested to values of $M_s \gg M_{u,d}$, we set
$M_{u,d} = 0$ in the quark loops.

To begin with we evaluate analytically the squared masses of the
excitations at the leading order in the parameter $M_s^2/\mu\Delta$.
The result is achieved evaluating ${\cal L}(p= 0)$ with ${\cal L}$
given by the sum of Eqs.~\eqref{eq:tadpoleMom}
and~\eqref{eq:selfMom}. We find
\begin{eqnarray}
{\cal L}(p=0) &=& \mu^2_{4} K^+ K^- + \mu^2_6 K^0 \bar{K}^0~,
\label{eq:LLL}
\end{eqnarray}
where the meson fields are defined in terms of the pion fields
$\pi_a$ as usual: $K^{\pm} = (\pi_4 \mp i \pi_5)/\sqrt{2}$, and
$K^0/\bar{K}^0 = (\pi_6 \mp i \pi_7)/\sqrt{2}$. In the above
equation
\begin{eqnarray}
\mu^2_{4} &=& \left(\frac{M_s^2}{2\mu}\right)^2~,\label{eq:Fkp}
\end{eqnarray}
\begin{eqnarray}
\mu^2_{6} &=& \mu^2_{4}~.\label{eq:Fk0}
\end{eqnarray}
For the other meson excitations we find a vanishing effective mass,
${\cal L}(p=0) = 0$.

\begin{figure*}[f]
\begin{center}
\includegraphics[width=14cm]{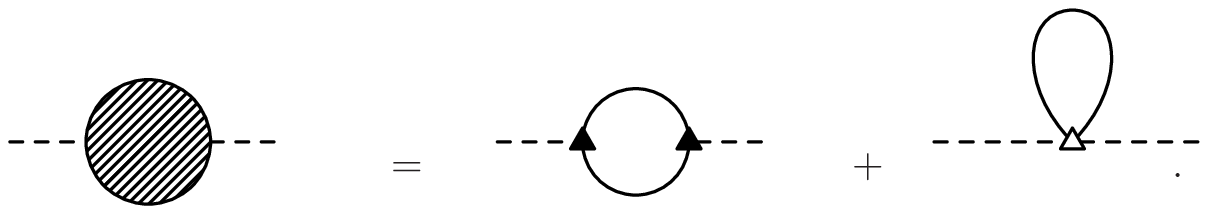}
\caption{\label{Fig:eff1}One loop effective action of the ``pion''
modes $\pi_a$. External dashed lines denote the meson field, while
solid lines stem for the quark propagators. Black triangles are for
$K_3$ vertices, while empty triangles are for $K_4$ vertices.
}\end{center}
\end{figure*}

The results~\eqref{eq:Fkp},~\eqref{eq:Fk0} are in agreement with
those of Bedaque and Sch\"{a}fer (BS) usually quoted in the
literature~\cite{Bedaque:2001je}. We now discuss how they are
obtained in the microscopic calculation. From now on we consider
only the charged kaons mode: the neutral ones are treated in a
similar way.

In Fig.~\ref{Fig:eff1} we draw the one loop effective action of
$\pi_4$ at zero external momentum (for $\pi_5$ one gets the same
results). External lines denote the meson field, solid line are
fermion propagators. We find
\begin{equation}
\parbox{20mm}{\includegraphics[width=3cm]{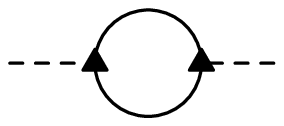}}
\quad\quad\quad=\frac{-i \pi_4^2}{8F^2}\frac{\mu^2}{4\pi^3}{\cal
K}(M_s)~,\label{eq:K}
\end{equation}
with ${\cal K}(M_s) = {\cal K}_0 + {\cal K}_2$ and
\begin{equation}
{\cal K}_0 = 16\Delta^2\int d^2\ell \frac{-\ell_0^2 +
3\Delta^2+\ell_\parallel^2}{(\ell_0^2 - \ell_\parallel^2 -\Delta^2)
(\ell_0^2 - \ell_\parallel^2 -4\Delta^2)}~,
\end{equation}
\begin{eqnarray}
{\cal K}_2 &=& -16\Delta^2
\left(\frac{M_s^2}{2\mu}\right)^2\nonumber\\
&&\times\int d^2\ell \frac{-\ell_0^4 +
\ell_0^2\Delta^2+3\Delta^4+4\Delta^2\ell_\parallel^2+\ell_\parallel^4}{(\ell_0^2
- \ell_\parallel^2 -\Delta^2)^3 (\ell_0^2 - \ell_\parallel^2
-4\Delta^2)}.
\end{eqnarray}
Moreover we find
\begin{equation}
\parbox{20mm}{\includegraphics[width=3cm]{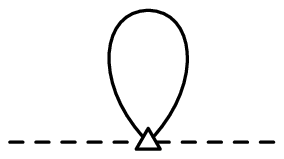}}
\quad\quad\quad=\frac{-i \pi_4^2}{8F^2}\frac{\mu^2}{4\pi^3}{\cal
N}(M_s)~,\label{eq:N}
\end{equation}
with ${\cal N}(M_s) = -{\cal K}_0 + {\cal N}_2(M_s)$ and
\begin{eqnarray}
{\cal N}_2 = 4\Delta^2 \left(\frac{M_s^2}{2\mu}\right)^2 \int
d^2\ell \frac{3\ell_0^2 + \Delta^2 + \ell_\parallel^2}{(\ell_0^2 -
\ell_\parallel^2 -\Delta^2)^3 }~.
\end{eqnarray}
\begin{widetext}
The ${\cal K}$ is canceled by the tadpole ${\cal N}$ if $M_s = 0$ as
it should, since for $M_s = 0$ the excitation has to be a true
Goldstone boson; adding the $\pi_5$ contribution the one loop
effective Lagrangian at zero momentum is therefore
\begin{eqnarray}
\parbox{20mm}{\includegraphics[width=4cm]{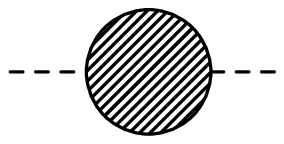}}
\quad\quad\quad~=~\frac{\pi_4^2 + \pi_5^2}{2}\frac{\mu^2}{\pi^2 F^2}
\frac{21-8\log2}{36} \left(\frac{M_s^2}{2\mu}\right)^2 \equiv
\mu_{4}^2 K^+ K^-~,\text{c.d.d.} \label{eq:comp}
\end{eqnarray}
where we have used the expression of $F$ in the CFL phase,
appropriated since we are working at the leading order in the
strange quark
mass~\cite{Bedaque:2001je,Son:1999cm,Casalbuoni:2000na},
\begin{equation}
F^2(M_s = 0) = \frac{\mu^2}{\pi^2}\frac{21-8\log2}{36}~.
\label{eq:F0}
\end{equation}
Eq.~\eqref{eq:comp} is in agreement
with~\cite{Bedaque:2001je,Schafer:2001za}; to our knowledge it is
the first time that this important result is deduced by a one loop
calculation of the effective action.

Before going on we comment on the choice of the chemical potentials.
We have repeated the above calculation setting $\mu_8$ to an
arbitrary value, and for simplicity leaving $\mu_e = \mu_3 = 0$. The
loop expressions in this case are complicated and therefore we do
not show them, but the final result is very simple, namely:
\begin{equation}
{\cal L}(p=0) = -\mu_8\frac{M_s^2}{2\mu}\left(K^+ K^- +
K^0\bar{K}^0\right)~.\label{eq:HiroSugg}
\end{equation}
We notice that the result is strongly dependent on the choice of
$\mu_8$. We obtain the correct result~\eqref{eq:comp}, that is the
squared mass consistent with the effective lagrangian
approach~\cite{Bedaque:2001je,Kryjevski:2003cu,Buballa:2004sx,Forbes:2004ww},
if and only if we set $\mu_8 = -M_s^2/2\mu$, its value in the
neutral CFL phase. As stressed in the introduction, this result is
not surprising: as a matter of fact, it has been shown
in~\cite{Kryjevski:2003cu,Forbes:2004ww} that in order to properly
write the effective lagrangian of the neutral CFL phase in terms of
the colorless pion fields, one has to get rid of the colored
components, and this is achieved if and only if the neutrality
conditions are properly implemented. Thus, in order to reproduce the
result of the effective lagrangian from the microscopic calculation,
one has to chose the chemical potentials in order to fulfill the
neutrality conditions.
\end{widetext}

As explained in the previous section, when one computes the meson
masses, one has to add  to the lagrangian in Eq.~\eqref{eq:LLL} the
shift of the vacuum energy due to finite quark masses, related in
the high effective theory to operators of order $1/\mu^2$. In the
CFL phase with massless quark propagators the result of this
calculation is~\cite{Son:1999cm,Beane:2000ms,Schafer:2001za}
\begin{equation}
m_{K^\pm}^2 = \frac{4A}{F^2}M_d(M_u + M_s)~,\label{eq:MK111}
\end{equation}
\begin{equation}
m_{K^0/\bar{K}^0}^2 = \frac{4A}{F^2}M_u(M_s + M_d)~,
\label{eq:MK0111}
\end{equation}
with $A = \Delta\bar\Delta/2\pi^2 \log(\mu/\Delta)$ in a NJL model
(in QCD $A = 3\Delta^2/4\pi^2$). As already explained in the
previous section, these results are not modified by the finite mass
and the color chemical potential in the quark loops. Moreover we
should notice that such kind of corrections vanish in the limit
$M_{u,d} \rightarrow 0$, which is the limit we are taking since in
the neutrality conditions in use we do not consider the effect of
the light quarks; on the other hand, the effective chemical
potential terms~\eqref{eq:Fkp},~\eqref{eq:Fk0} survive in this
limiting case. For these reasons we do no longer discuss the
contribution $m_K$ arising from to the anti-gap action, using for
them the leading order results~\eqref{eq:MK111},~\eqref{eq:MK0111}.
As for the charged pions and the unflavored meson fields, at this
order they are massless.

The complete lagrangian of the kaon modes at $p=0$ is thus given by
\begin{equation}
{\cal L}(p=0) = (\mu_4^2 - m_{K^\pm}^2) K^+ K_- +
       (\mu_6^2 - m_{K^0}^2) K^0 \bar{K}_0~.
\end{equation}
The terms~\eqref{eq:Fkp},~\eqref{eq:Fk0} are usually referred as
effective chemical potentials, as they enter into the meson
lagrangian via a covariant derivative as a typical meson chemical
potential does,
\begin{equation}
{\cal L} \thicksim [(\partial_0 + i\mu_4)K^+][(\partial_0 -
i\mu_4)K^-]~.
\end{equation}
Since they give rise to a mass term with the wrong sign, when $\mu_4
> m_K$ the ground state is unstable toward the formation
of a kaon condensate~\cite{Bedaque:2001je}. In our approximation
scheme, $m_K \simeq 0$ and $\mu_4^2 >0$, thus the kaon condensation
occurs for each value of $M_s \neq 0$.

\subsection{Numerical results}
The above results~\eqref{eq:Fkp},~\eqref{eq:Fk0} are confirmed by a
numerical evaluation of the one loop
diagrams~\eqref{eq:tadpoleMom},~\eqref{eq:selfMom} for values of
$M_s^2/\mu$ up to $2\Delta$; above this critical value the
transition to gCFL phase occurs and one has to consider the
contribution of the gapless modes in the quark loops, together with
the conditions $\mu_e \neq 0$ and $\mu_3 \neq 0$, which is beyond
the scope of this paper.

\begin{figure*}[f] \begin{center}
\includegraphics[width=14cm]{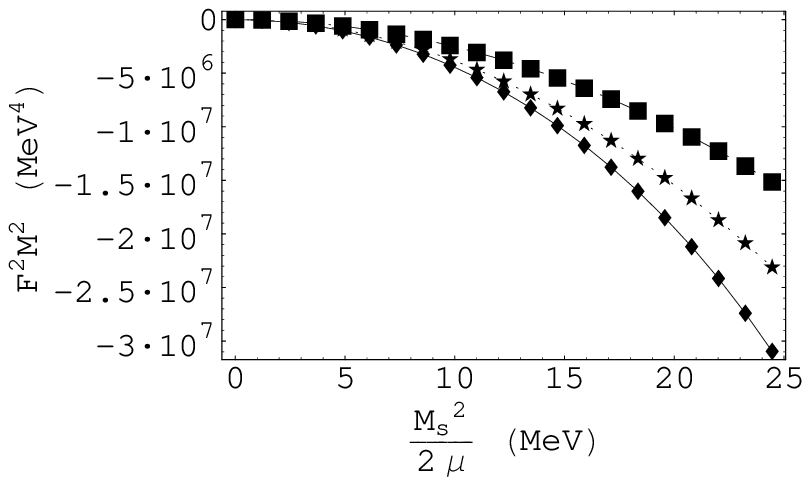}
\end{center}
\caption{\label{Fig:Mu4}Squared mass of the kaon modes times $F^2$,
in the neutral CFL phase with massless $u$ and $d$ quarks, against
$M_s^2/2\mu$. Diamonds and stars correspond respectively to $\Delta
= 25$ MeV and $\Delta = 75$ MeV. Squares correspond to the leading
order solution, which does not depend on $\Delta$. In the figure we
have chosen $\mu = 500$ MeV as a reference value of the baryon
chemical potential.}
\end{figure*}
\begin{widetext}
The result of such a calculation is shown in Fig.~\ref{Fig:Mu4},
where  we plot $F^2$ times the squared mass of the kaon modes $M^2
\equiv -\mu_4^2$, obtained by the numerical evaluation of the loops
in the case $M_u = M_d = 0$ (in this case the kaon mass coincides
with the minus of the effective chemical potential $\mu_4^2$ since
$m_K = 0$). In the figure, diamonds and stars correspond
respectively to $\Delta = 25$ MeV and $\Delta = 75$ MeV; the squares
correspond to the leading order solution, which does not depend on
$\Delta$. The dependence on the gap parameter $\Delta$ when one
approaches the value $M_s^2/2\mu\approx\Delta$ can be easily
understood by looking at the next-to-leading order correction to the
result quoted in Eq.~\eqref{eq:comp}. We find
\begin{eqnarray}
&&
F^2\times\!\!\!\!\!\!\parbox{20mm}{\includegraphics[width=4cm]{Loop4.eps}}
\quad\quad\quad~=~K^+ K^-\left(\frac{M_s^2}{2\mu}\right)^2\left[c_0
+ c_2 \left(\frac{M_s^2}{2\mu\Delta}\right)^2\right]~,
  \label{eq:comp2}
\end{eqnarray}
where $c_0 = (21-8\log2)\mu^2/36\pi^2$ as can be read from
Eq.~\eqref{eq:comp} and $c_2 \approx 0.07$: the introduction of the
higher order corrections introduces a dependence on $\Delta$ that
vanishes only when $M_s^2/2\mu\Delta\rightarrow0$.
\end{widetext}

We notice that at a given $M_s$, the inclusion of the higher order
corrections increases the value of $\mu_4$, the effect being more
important for weak couplings. On the other hand, once $M_{u,d} \neq
0$, these corrections do not change $m_K$~\eqref{eq:MK111} in a
significant way: as a matter of fact the relevant diagrams get the
leading contribution from the hard region $\ell_\parallel \simeq\mu
\gg \mu_8,~M_s$ and thus do not depend on $\mu_8,~M_s$. Therefore at
a given $M_s$ the higher order corrections have the net effect of
increasing $\mu_4 - m_K$ with respect to the leading order result,
thus favoring kaon condensation.

\section{Conclusions}
In this paper we have computed the effective chemical potentials of
the $SU(3)_A$ Goldstone excitations in the neutral CFL phase of QCD,
starting from a microscopic model of quarks interacting with the
Goldstones. We have worked within the approximations $M_u = M_d =0$,
thus retaining the neutrality conditions $\mu_e = \mu_3 = 0$, $\mu_8
\approx -M_s^2/2\mu$. Our results agree with those of Bedaque and
Schafer~\cite{Bedaque:2001je}, which are obtained by the authors on
the basis of symmetry arguments and not starting from a microscopic
model; therefore we offer here a diagrammatic derivation of their
results, showing how they arise from loop effects.

In addition to the new derivation of the classical results, we have
performed a numerical evaluation of the low energy parameters of the
effective lagrangian of the meson modes, valid in the whole CFL
domain $M_s^2/2\mu\Delta \leqslant 1$. Our results, summarized in
Fig~\ref{Fig:Mu4}, show an enhancement of the kaon condensation with
respect to the leading order result usually referred to in the
literature.

We have performed the calculation in the color neutral state: this
choice is motivated by gauge invariance, which requires that any
homogeneous ground state of QCD has to be color
neutral~\cite{Kryjevski:2003cu,Forbes:2004ww}. Moreover, this is the
only way to properly reproduce the low energy effective action of
Ref.~\cite{Bedaque:2001je} starting from the microscopic theory.
Indeed it has been shown~\cite{Kryjevski:2003cu,Forbes:2004ww} that
the requirement of color neutrality is equivalent to the removal,
from the low energy effective lagrangian, of the color non-singlet
fields. Translated to the microscopic calculation language, this is
the same to say that the quark chemical potentials {\em must} be
chosen in order to satisfy the color neutrality conditions. This is
a necessary requirement of any calculation of the properties of the
Goldstone bosons: any violation of color neutrality in the
microscopic model is translated, in the effective theory, into the
presence of non-singlet fields in the low-energy lagrangian.   This
reasoning is reinforced by the calculation leading to
Eq.~\eqref{eq:HiroSugg}, where we have set $\mu_8$ to an arbitrary
value: we obtain the correct result, consistent with the low energy
effective lagrangian, if and only if $\mu_8$ is equal to its value
in the neutral phase.

We have not computed both $F$ and the velocity $v$ of the meson
fields. Their values in the neutral phase can be obtained from the
results quoted in Refs.~\cite{Casalbuoni:2004tb} once we notice that
in the gapped CFL phase~\cite{Son:1999cm}
\begin{equation}
F^2_a \propto m_{D,a}^2~,~~~~~v^2_a = m_{M,a}^2/m_{D,a}^2~,
\end{equation}
where $F_a$ is the decay constant of the field $\pi_a$, and
$m_{D,a}$ ($m_{M,a}$) correspond to the Debye (Meissner) screening
mass of the gluon with adjoint index $a$.

It would be interesting to extend our computations to the case of
$M_{u,d} \neq 0$, in order to reproduce the classical results
of~\cite{Bedaque:2001je} for non vanishing light quark masses. To do
that one has to consider the effect of the light quark masses to the
neutrality conditions in order to make a consistent microscopic
calculation. We expect that these effects are important in the
region $M_s^2/\mu\Delta \ll 1$, becoming less and less important
when one approaches the onset of the CFL$\rightarrow$gCFL transition
where $M_s \gg M_{u,d}$. Moreover, also retaining the approximation
$M_u = M_d = 0$, it would be fine to increase the strange quark mass
beyond the onset CFL$\rightarrow$gCFL, in order to understand the
role of the gapless fermion excitations on the spectrum of the axial
excitations. From the formal point of view it is sufficient to
introduce the proper values of the charge and color chemical
potentials, but a preliminary analysis shows that the numerical work
required for this interesting project is much more involved than the
one presented here. Beside masses, the effective lagrangian
parameters $F$ and $v$ should be evaluated in the gCFL regime: from
the two flavor case we know that the squared velocities of some of
the fields can be negative~\cite{Gatto:2007ja}, and this would lead
to the Goldstone currents studied in~\cite{Schafer:2005ym}. Finally,
last but not least, it would be interesting to extend the
calculations to finite temperature. Unfortunately this project is
not trivial from the numerical point of view as it requires the
introduction of $\mu_e$, $\mu_3$ beside
$\mu_8$~\cite{Alford:2003fq}. We leave also this point to a future
project.

\acknowledgments We thank M.~Ciminale, N.~Ippolito, and T.~Schafer
for enlightening discussions and critical comments. We thank
R.~Gatto and G.~Nardulli for a careful reading of the manuscript.
Moreover we thank H.~Abuki for having suggested the calculations
leading to Eq.~\eqref{eq:HiroSugg}.


\newpage

\begin{thebibliography}{99}

\bibitem{Abuki:2004zk}
  H.~Abuki, M.~Kitazawa and T.~Kunihiro,
  Phys.\ Lett.\  B {\bf 615}, 102 (2005)
  [arXiv:hep-ph/0412382];
  H.~Abuki and T.~Kunihiro,
  Nucl.\ Phys.\  A {\bf 768}, 118 (2006)
  [arXiv:hep-ph/0509172].

\bibitem{Ruster:2005jc}
  S.~B.~Ruster, V.~Werth, M.~Buballa, I.~A.~Shovkovy and D.~H.~Rischke,
  Phys.\ Rev.\  D {\bf 72}, 034004 (2005)
  [arXiv:hep-ph/0503184].

\bibitem{Blaschke:2005uj}
  D.~Blaschke, S.~Fredriksson, H.~Grigorian, A.~M.~Oztas and F.~Sandin,
  Phys.\ Rev.\  D {\bf 72}, 065020 (2005)
  [arXiv:hep-ph/0503194].

\bibitem{Hatsuda:2006ps}
  T.~Hatsuda, M.~Tachibana, N.~Yamamoto and G.~Baym,
  Phys.\ Rev.\ Lett.\  {\bf 97}, 122001 (2006)
  [arXiv:hep-ph/0605018].

\bibitem{Yamamoto:2007ah}
  N.~Yamamoto, M.~Tachibana, T.~Hatsuda and G.~Baym,
  arXiv:0704.2654 [hep-ph].

\bibitem{Rajagopal:2000wf}
  K.~Rajagopal and F.~Wilczek,
  arXiv:hep-ph/0011333.

\bibitem{Alford:2001dt}
  M.~G.~Alford,
  Ann.\ Rev.\ Nucl.\ Part.\ Sci.\  {\bf 51}, 131 (2001)
  [arXiv:hep-ph/0102047].

\bibitem{Nardulli:2002ma}
  G.~Nardulli,
  Riv.\ Nuovo Cim.\  {\bf 25N3}, 1 (2002)
  [arXiv:hep-ph/0202037].

\bibitem{Reddy:2002ri}
  S.~Reddy,
  Acta Phys.\ Polon.\  B {\bf 33}, 4101 (2002)
  [arXiv:nucl-th/0211045].

\bibitem{Schafer:2003vz}
  T.~Schafer,
  arXiv:hep-ph/0304281.

\bibitem{Rischke:2003mt}
  D.~H.~Rischke,
  Prog.\ Part.\ Nucl.\ Phys.\  {\bf 52}, 197 (2004)
  [arXiv:nucl-th/0305030].

\bibitem{Buballa:2003qv}
  M.~Buballa,
  Phys.\ Rept.\  {\bf 407}, 205 (2005)
  [arXiv:hep-ph/0402234].

\bibitem{Huang:2004ik}
  M.~Huang,
  Int.\ J.\ Mod.\ Phys.\  E {\bf 14}, 675 (2005)
  [arXiv:hep-ph/0409167].

\bibitem{Shovkovy:2004me}
  I.~A.~Shovkovy,
  Found.\ Phys.\  {\bf 35}, 1309 (2005)
  [arXiv:nucl-th/0410091].

\bibitem{Alford:2006fw}
  M.~Alford and K.~Rajagopal,
  arXiv:hep-ph/0606157.

\bibitem{Alford:1998mk}
  M.~G.~Alford, K.~Rajagopal and F.~Wilczek,
  Nucl.\ Phys.\  B {\bf 537}, 443 (1999)
  [arXiv:hep-ph/9804403].

\bibitem{Evans:1999at}
  N.~J.~Evans, J.~Hormuzdiar, S.~D.~H.~Hsu and M.~Schwetz,
  Nucl.\ Phys.\  B {\bf 581}, 391 (2000)
  [arXiv:hep-ph/9910313].

\bibitem{Alford:2003fq}
  M.~Alford, C.~Kouvaris and K.~Rajagopal,
  Phys.\ Rev.\ Lett.\  {\bf 92}, 222001 (2004)
  [arXiv:hep-ph/0311286];
  M.~Alford, C.~Kouvaris and K.~Rajagopal,
  Phys.\ Rev.\ D {\bf 71}, 054009 (2005)
  [arXiv:hep-ph/0406137];
  K.~Fukushima, C.~Kouvaris and K.~Rajagopal,
  Phys.\ Rev.\ D {\bf 71}, 034002 (2005)
  [arXiv:hep-ph/0408322].

\bibitem{Bedaque:2001je}
  P.~F.~Bedaque and T.~Schafer,
  Nucl.\ Phys.\  A {\bf 697}, 802 (2002)
  [arXiv:hep-ph/0105150];
  P.~F.~Bedaque,
  Phys.\ Lett.\  B {\bf 524}, 137 (2002)
  [arXiv:nucl-th/0110049];
  T.~Schafer,
  Phys.\ Rev.\ Lett.\  {\bf 85}, 5531 (2000)
  [arXiv:nucl-th/0007021].

\bibitem{Kaplan:2001qk}
  D.~B.~Kaplan and S.~Reddy,
  Phys.\ Rev.\  D {\bf 65}, 054042 (2002)
  [arXiv:hep-ph/0107265].

\bibitem{Casalbuoni:2005zp}
  R.~Casalbuoni, R.~Gatto, N.~Ippolito, G.~Nardulli and M.~Ruggieri,
  Phys.\ Lett.\  B {\bf 627}, 89 (2005)
  [Erratum-ibid.\  B {\bf 634}, 565 (2006)]
  [arXiv:hep-ph/0507247];
  M.~Mannarelli, K.~Rajagopal and R.~Sharma,
  Phys.\ Rev.\  D {\bf 73}, 114012 (2006)
  [arXiv:hep-ph/0603076];
  K.~Rajagopal and R.~Sharma,
  Phys.\ Rev.\  D {\bf 74}, 094019 (2006)
  [arXiv:hep-ph/0605316];
  N.~D.~Ippolito, G.~Nardulli and M.~Ruggieri,
  JHEP {\bf 0704}, 036 (2007)
  [arXiv:hep-ph/0701113].

\bibitem{Shovkovy:2003uu}
  I.~Shovkovy and M.~Huang,
  Phys.\ Lett.\  B {\bf 564}, 205 (2003)
  [arXiv:hep-ph/0302142];
  M.~Huang and I.~Shovkovy,
  Nucl.\ Phys.\  A {\bf 729}, 835 (2003)
  [arXiv:hep-ph/0307273].

\bibitem{Gubankova:2003uj}
  E.~Gubankova, W.~V.~Liu and F.~Wilczek,
  Phys.\ Rev.\ Lett.\  {\bf 91}, 032001 (2003)
  [arXiv:hep-ph/0304016].

\bibitem{Alford:2000ze}
  M.~G.~Alford, J.~A.~Bowers and K.~Rajagopal,
  Phys.\ Rev.\  D {\bf 63}, 074016 (2001)
  [arXiv:hep-ph/0008208];
  J.~A.~Bowers and K.~Rajagopal,
  Phys.\ Rev.\  D {\bf 66}, 065002 (2002)
  [arXiv:hep-ph/0204079];
  R.~Casalbuoni, M.~Ciminale, M.~Mannarelli, G.~Nardulli, M.~Ruggieri and R.~Gatto,
  Phys.\ Rev.\  D {\bf 70}, 054004 (2004)
  [arXiv:hep-ph/0404090].

\bibitem{Schafer:2000tw}
  T.~Schafer,
  Phys.\ Rev.\  D {\bf 62}, 094007 (2000)
  [arXiv:hep-ph/0006034];
  M.~G.~Alford, J.~A.~Bowers, J.~M.~Cheyne and G.~A.~Cowan,
  Phys.\ Rev.\  D {\bf 67}, 054018 (2003)
  [arXiv:hep-ph/0210106];
  A.~Schmitt,
  Phys.\ Rev.\  D {\bf 71}, 054016 (2005)
  [arXiv:nucl-th/0412033].

\bibitem{Aguilera:2005tg}
  D.~N.~Aguilera, D.~Blaschke, M.~Buballa and V.~L.~Yudichev,
  Phys.\ Rev.\  D {\bf 72}, 034008 (2005)
  [arXiv:hep-ph/0503288];
  F.~Marhauser, D.~Nickel, M.~Buballa and J.~Wambach,
  Phys.\ Rev.\  D {\bf 75}, 054022 (2007)
  [arXiv:hep-ph/0612027].

\bibitem{Huang:2004bg}
  M.~Huang and I.~A.~Shovkovy,
  Phys.\ Rev.\  D {\bf 70}, 051501 (2004)
  [arXiv:hep-ph/0407049];
  M.~Huang and I.~A.~Shovkovy,
  Phys.\ Rev.\  D {\bf 70}, 094030 (2004)
  [arXiv:hep-ph/0408268].

\bibitem{Casalbuoni:2004tb}
  R.~Casalbuoni, R.~Gatto, M.~Mannarelli, G.~Nardulli and M.~Ruggieri,
  Phys.\ Lett.\  B {\bf 605}, 362 (2005)
  [Erratum-ibid.\  B {\bf 615}, 297 (2005)]
  [arXiv:hep-ph/0410401];
  M.~Alford and Q.~h.~Wang,
  J.\ Phys.\ G {\bf 31}, 719 (2005)
  [arXiv:hep-ph/0501078];
  K.~Fukushima,
  Phys.\ Rev.\  D {\bf 72}, 074002 (2005)
  [arXiv:hep-ph/0506080].

\bibitem{Gorbar:2005rx}
  E.~V.~Gorbar, M.~Hashimoto and V.~A.~Miransky,
  Phys.\ Lett.\  B {\bf 632}, 305 (2006)
  [arXiv:hep-ph/0507303];
  E.~V.~Gorbar, M.~Hashimoto and V.~A.~Miransky,
  Phys.\ Rev.\  D {\bf 75}, 085012 (2007)
  [arXiv:hep-ph/0701211];
  M.~Hashimoto and V.~A.~Miransky,
  arXiv:0705.2399 [hep-ph].

\bibitem{Huang:2005pv}
  M.~Huang,
  Phys.\ Rev.\  D {\bf 73}, 045007 (2006)
  [arXiv:hep-ph/0504235].

\bibitem{Gatto:2007ja}
  R.~Gatto and M.~Ruggieri,
  arXiv:hep-ph/0703276.

\bibitem{Schafer:2005ym}
  T.~Schafer,
  Phys.\ Rev.\ Lett.\  {\bf 96}, 012305 (2006)
  [arXiv:hep-ph/0508190];
  A.~Gerhold and T.~Schafer,
  Phys.\ Rev.\  D {\bf 73}, 125022 (2006)
  [arXiv:hep-ph/0603257];
  A.~Gerhold, T.~Schafer and A.~Kryjevski,
  Phys.\ Rev.\  D {\bf 75}, 054012 (2007)
  [arXiv:hep-ph/0612181].

\bibitem{Giannakis:2004pf}
  I.~Giannakis and H.~C.~Ren,
  Phys.\ Lett.\  B {\bf 611}, 137 (2005)
  [arXiv:hep-ph/0412015];
\bibitem{Giannakis:2005vw}
  I.~Giannakis and H.~C.~Ren,
  Nucl.\ Phys.\  B {\bf 723}, 255 (2005)
  [arXiv:hep-th/0504053];
  I.~Giannakis, D.~Hou, M.~Huang and H.~c.~Ren,
  Phys.\ Rev.\  D {\bf 75}, 011501 (2007)
  [arXiv:hep-ph/0606178];
  M.~Ciminale, G.~Nardulli, M.~Ruggieri and R.~Gatto,
  Phys.\ Lett.\  B {\bf 636}, 317 (2006)
  [arXiv:hep-ph/0602180].

\bibitem{Hong:1998tn}
  D.~K.~Hong,
  Phys.\ Lett.\  B {\bf 473}, 118 (2000)
  [arXiv:hep-ph/9812510];
  D.~K.~Hong,
  Nucl.\ Phys.\  B {\bf 582}, 451 (2000)
  [arXiv:hep-ph/9905523].

\bibitem{Schafer:2003jn}
  T.~Schafer,
  Nucl.\ Phys.\  A {\bf 728}, 251 (2003)
  [arXiv:hep-ph/0307074].

\bibitem{Casalbuoni:1999wu}
  R.~Casalbuoni and R.~Gatto,
  Phys.\ Lett.\  B {\bf 464}, 111 (1999)
  [arXiv:hep-ph/9908227].

\bibitem{Kaplan:2001hh}
  D.~B.~Kaplan and S.~Reddy,
  Phys.\ Rev.\ Lett.\  {\bf 88}, 132302 (2002)
  [arXiv:hep-ph/0109256];
  P.~Jaikumar, M.~Prakash and T.~Schafer,
  Phys.\ Rev.\  D {\bf 66}, 063003 (2002)
  [arXiv:astro-ph/0203088];
  M.~G.~Alford, M.~Braby, S.~Reddy and T.~Schafer,
  arXiv:nucl-th/0701067.

\bibitem{Son:1999cm}
  D.~T.~Son and M.~A.~Stephanov,
  Phys.\ Rev.\ D {\bf 61}, 074012 (2000)
  [arXiv:hep-ph/9910491];
  D.~T.~Son and M.~A.~Stephanov,
  Phys.\ Rev.\ D {\bf 62}, 059902 (2000)
  [arXiv:hep-ph/0004095].

\bibitem{Casalbuoni:2000na}
  R.~Casalbuoni, R.~Gatto and G.~Nardulli,
  Phys.\ Lett.\ B {\bf 498}, 179 (2001)
  [Erratum-ibid.\ B {\bf 517}, 483 (2001)]
  [arXiv:hep-ph/0010321].


\bibitem{Beane:2000ms}
  S.~R.~Beane, P.~F.~Bedaque and M.~J.~Savage,
  Phys.\ Lett.\ B {\bf 483}, 131 (2000)
  [arXiv:hep-ph/0002209].


\bibitem{Schafer:2001za}
  T.~Schafer,
  Phys.\ Rev.\ D {\bf 65}, 074006 (2002)
  [arXiv:hep-ph/0109052].

\bibitem{Manuel:2000wm}
  C.~Manuel and M.~H.~G.~Tytgat,
  Phys.\ Lett.\ B {\bf 479}, 190 (2000)
  [arXiv:hep-ph/0001095].

\bibitem{Kryjevski:2003cu}
  A.~Kryjevski,
  Phys.\ Rev.\  D {\bf 68}, 074008 (2003)
  [arXiv:hep-ph/0305173].

\bibitem{Buballa:2004sx}
  M.~Buballa,
  Phys.\ Lett.\ B {\bf 609}, 57 (2005)
  [arXiv:hep-ph/0410397].

\bibitem{Forbes:2004ww}
  M.~M.~Forbes,
  Phys.\ Rev.\  D {\bf 72}, 094032 (2005)
  [arXiv:hep-ph/0411001].

\bibitem{Ebert:2006tc}
  D.~Ebert and K.~G.~Klimenko,
  Phys.\ Rev.\  D {\bf 75}, 045005 (2007)
  [arXiv:hep-ph/0611385];
  D.~Ebert, K.~G.~Klimenko and V.~L.~Yudichev,
  arXiv:0705.2666 [hep-ph].

\bibitem{Eguchi:1976iz}
  T.~Eguchi,
  Phys.\ Rev.\  D {\bf 14}, 2755 (1976).

\bibitem{Anglani:2007aa}
  R.~Anglani, R.~Gatto, N.~D.~Ippolito, G.~Nardulli and M.~Ruggieri,
  arXiv:0706.1781 [hep-ph].

\bibitem{Alford:2002kj}
  M.~Alford and K.~Rajagopal,
  JHEP {\bf 0206}, 031 (2002)
  [arXiv:hep-ph/0204001].

\bibitem{Nambu:1961tp}
  Y.~Nambu and G.~Jona-Lasinio,
  Phys.\ Rev.\  {\bf 122}, 345 (1961);
  Y.~Nambu and G.~Jona-Lasinio,
  Phys.\ Rev.\  {\bf 124}, 246 (1961).


\bibitem{Klevansky:1992qe}
  S.~P.~Klevansky,
  Rev.\ Mod.\ Phys.\  {\bf 64} (1992) 649.
\bibitem{Hatsuda:1994pi}
  T.~Hatsuda and T.~Kunihiro,
  Phys.\ Rept.\  {\bf 247}, 221 (1994)
  [arXiv:hep-ph/9401310].

\bibitem{Buballa:2005bv}
  M.~Buballa and I.~A.~Shovkovy,
  Phys.\ Rev.\ D {\bf 72}, 097501 (2005)
  [arXiv:hep-ph/0508197].



\bibitem{Casalbuoni:2002st}
  R.~Casalbuoni, F.~De Fazio, R.~Gatto, G.~Nardulli and M.~Ruggieri,
  Phys.\ Lett.\  B {\bf 547}, 229 (2002)
  [arXiv:hep-ph/0209105].

\bibitem{Steiner:2002gx}
  A.~W.~Steiner, S.~Reddy and M.~Prakash,
  Phys.\ Rev.\  D {\bf 66}, 094007 (2002)
  [arXiv:hep-ph/0205201].

\end{thebibliography}
\end{document}